\documentclass{kluwer}    % Specifies the document style.
\usepackage[]{graphicx}

\newdisplay{guess}{Conjecture}

\begin{document}
\begin{article}
\begin{opening}         

\title{Cepheid Limb Darkening Models for the VLTI}

\author{Massimo \surname{Marengo}}
\author{Margarita \surname{Karovska}}
\author{Dimitar D. \surname{Sasselov}}
\author{Costas \surname{Papaliolios}}
\institute{Harvard-Smithsonian Center for Astrophysics, Cambridge, MA, USA}
\author{J.T. \surname{Armstrong}}
\institute{Remote Sensing Division, Naval Research Laboratory,
Washington, DC, USA} 
\author{Tyler E. \surname{Nordgren}}
\institute{Dept. of Physics. University of Redlands, Redlands, CA, USA}
\runningauthor{M. Marengo et al.}
\runningtitle{Cepheids Limb Darkening Models for the VLTI}
%\date{November 15, 2002}

\begin{abstract}
We present a new method to compute wavelength- and phase-dependent
limb darkening corrections for classical Cepheids. These corrections
are derived from hydrodynamic simulations and radiative transfer
modeling with a full set of atomic and molecular opacities. Comparison
with hydrostatic models having the same stellar parameters show a
larger limb darkening for most phases in our models, and temporal
variations related to the hydrodynamics of the stellar pulsation. We
assess the implications of our results with respect to the geometric
Baade-Wesselink method, which uses interferometric measurements of
Cepheid angular diameters to determine their distances and radii. The
relevance of the hydrodynamic effects predicted by our model on the
limb darkening of pulsating Cepheids is finally discussed in terms of
the predicted capabilities of the VLTI. 
\end{abstract}

%\keywords{Cepheids --- stars:atmosperes --- stars: oscillations ---
%techniques: interferometric}

\end{opening}

% -----------------------------------------------------------------------------
% Main paper section
% -----------------------------------------------------------------------------

\section{Introduction}\label{sec-intro}

The period-luminosity (PL) relation of Classical Cepheids, since its
discovery by \inlinecite{leavitt1906}, has played an important role
in setting the extragalactic distance scale. The reliability of this
relation, however, depends on the accurate calibration of its
zero-point. This is difficult to accomplish, due to the low density
distribution of Cepheids in our Galaxy. The determination of the PL
relation zero point from geometric parallaxes has been attempted using
the database of the Hipparcos mission \cite{feast1997}, but large
uncertainties do not allow an unambiguous interpretation of the
results \cite{madore1998}. The same is also true for the classical
Baade-Wesselink (BW) method (\opencite{baade1926};
\opencite{wesselink1946}), which suffers from large uncertainties
related to the source colors and metallicity.

The recent advances in interferometry, however, have finally allowed
the direct detection of pulsations in nearby Cepheids
(\opencite{lane2000}; \opencite{kervella2001};
\opencite{lane2002}). This result have opened the way for a
\emph{geometric} variant of the BW method \cite{sasselov1994},
allowing an independent distance measurement of pulsating stars, based
on the interferometric determination of their radial pulsations.

As pointed-out by \inlinecite{sasselov1994}, the feasibility of the
geometric BW method requires accurate predictions for the limb
darkening (LD) of the pulsating star, necessary to convert the
measured interferometric visibilities into angular diameters. In this
paper we present our time- and wavelength-dependent LD models, based
on hydrodynamic models of the classical Cepheids, and we discuss their
importance for future VLTI observations of this type of pulsating
stars.

% -----------------------------------------------------------------------------

\section{Dynamic models for Pulsating Cepheids}\label{sec-models}

In \inlinecite{marengo2002} we presented a new method for computing
accurate time- and wavelength-dependent center-to-limb brightness
distributions for classical Cepheids. This method is based on
hydrodynamic simulations of the Cepheid atmosphere, performed in
non-LTE conditions with a simplified radiative transfer
\cite{sasselov1994b}. The pulsations are induced by perturbing the
model atmosphere with a piston, parametrized by matching the radial
velocity curve, phase lags and other stellar observables. On top of
the hydrodynamic atmosphere, we solve the full LTE radiative transfer
problem, using the complete set of atomic and molecular opacities
provided by the ATLAS/SYNTHE program (\opencite{kurucz1970},
\citeyear{kurucz1979}, \citeyear{kurucz1993}). From the SYNTHE
emergent spectral intensity, we derive the LD of the pulsating star at
a number of wavelengths and pulsational phases.

The center-to-limb intensities from which the LD is derived are
computed with SYNTHE for a model atmosphere having (1) the effective
temperature $T_{eff}(\phi)$ obtained as a function of the pulsational
phase from observations \cite{krockenberger1997}, (2) $\log g$,
$L/H$, metallicity, $v_{turb}$ and the atmospheric Rosseland opacity
from a hydrostatic model approximating the hydrodynamic atmosphere at
each pulsational phase, and (3) $T(r)$, $n_e(r)$ and $P(r)$ from the
hydrodynamic model, excluding the chromospheric region.
The main limitation of this procedure resides in the
opacities being computed with a hydrostatic atmosphere in LTE
conditions. However, the resulting limb darkened intensities are still
much improved with respect to competing models, usually derived for
static yellow supergiants, since the thermodynamic state of the
atmosphere is completely set by the hydrodynamics.

\begin{figure}
\centerline{\includegraphics[width=15pc,angle=-90]{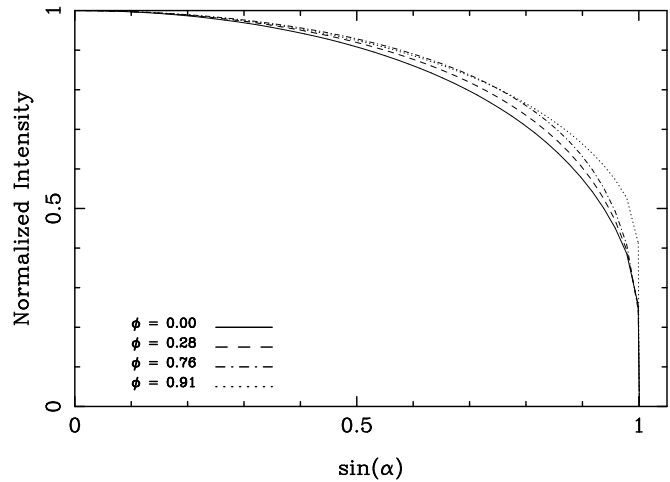}}
\caption{Limb-to-center profiles for $\zeta$~Gem at dynamical 
phases 0 (minimum radius), 0.28 (maximum luminosity), 0.76 (minimum
luminosity) and 0.91 (passage of a shockwave through the
photosphere). The profiles are computed for $\lambda =
500$~nm. The x-axis coordinate $\sin(\alpha)$ is the projected radial
coordinate of the star.}\label{fig1} 
\end{figure}

The radial profile of the hydrodynamic atmosphere can be very
different from the structure of a hydrostatic atmosphere with
equivalent $T_{eff}$ and $\log g$. This is especially true for
pulsational phases characterized by the propagation of
hydrodynamic instabilities or shocks, opposed to other phases where
the structure of the pulsating atmosphere is quasi-static. The fact
that the atmosphere structure changes as the Cepheid pulsates,
leads to different amount of limb darkening for different pulsational
phases. Figure~\ref{fig1} shows how the photosphere of the classical
Cepheid $\zeta$~Gem appears to be more limb darkened near the
minimum radius, and less limb darkened as a consequence of a shockwave
perturbing its atmosphere. 

The center-to-limb intensities shown in Figure~\ref{fig1} can be used
to derive the LD correction $k(\lambda,\phi)$ for the modeled
Cepheid. This correction is defined as the ratio between the
angular diameter measured by fitting the interferometer visibilities
with a uniform brightness disk (UD) model, with the \emph{true}
diameter of the limb darkened star (the ``limb darkened diameter''):
$k(\lambda,\phi) = \theta_{UD}(\lambda,\phi)/\theta_{LD}(\phi)$. The
LD correction is wavelength and phase-dependent, as a consequence of
the wavelength dependence of the atmospheric opacities, and because of
the changes in the atmospheric structure as the star pulsates. 

\begin{figure}
\centerline{\includegraphics[height=17pc]{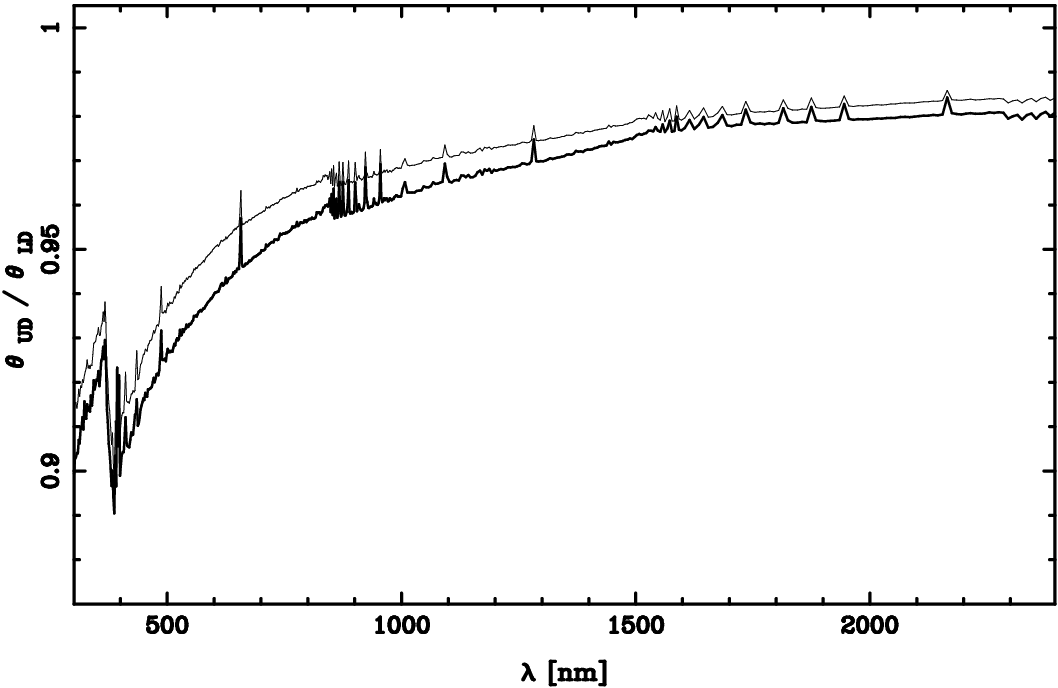}}
\caption{Wavelength dependence of the LD correction $k(\lambda)$ for
$\zeta$~Gem at minimum radius. The solid thick line is $k(\lambda)$ from
our hydrodynamic model; the thin line is derived from a
hydrostatic model having $T_{eff}$ and $\log g$ from
observations.}\label{fig2}
\end{figure}

\begin{figure}
\centerline{\includegraphics[width=18pc]{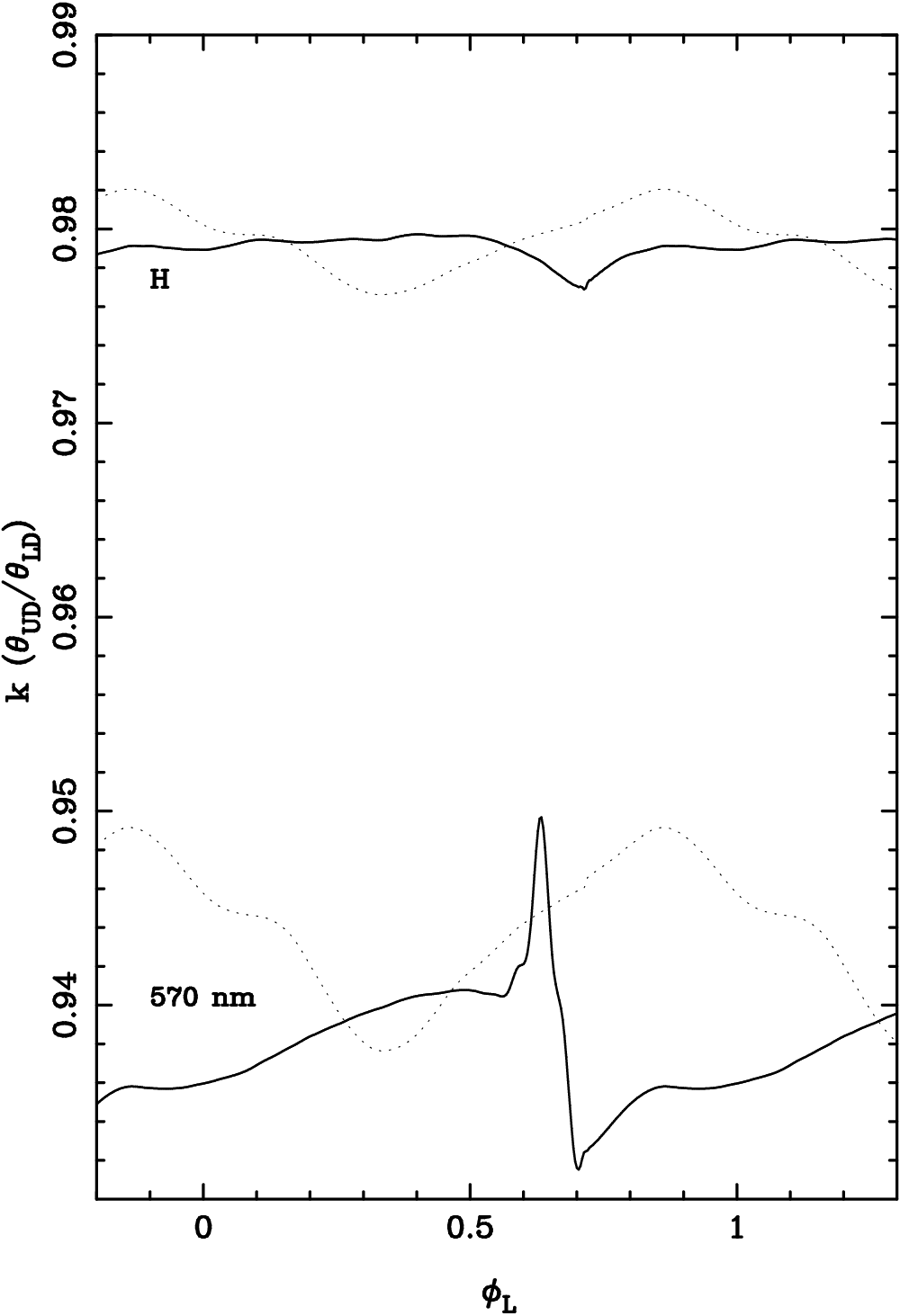}}
\caption{Phase dependent $k(\phi_L)$ for $\zeta$~Gem in the near-IR
H-band and in the visible. The solid line is $k(\phi_L)$ from our
hydrodynamic model, while the dotted line is derived from a
hydrostatic model having $T_{eff}$ and $\log g$ from
observations.}\label{fig3} 
\end{figure}

As an example, Figure~\ref{fig2} shows the spectral dependence of the
LD correction for our model of $\zeta$~Gem at minimum radius. The
limb darkening is larger at UV and optical wavelengths than in the
infrared. Neglecting to correct for LD in the near-IR,
however, will still produce an error of the order of $\sim 2$\% with
respect to the UD diameter of the star. Note that the LD correction
$k$ is very sensitive to the photospheric spectral lines, which can
thus be used as an effective diagnostic tool to better constrain the
LD with interferometric observations.

Figure~\ref{fig3} shows the variations of the LD corrections
with the pulsational phase, in the visible and at near-IR wavelengths
(H-band). The dotted lines are the LD corrections for of equivalent
hydrostatic atmosphere, having the same $T_{eff}$ and $\log g$ than
the hydrodynamic models. The hydrostatic LD corrections closely
follows the changes of $T_{eff}(\phi)$ with the pulsational phase. In
the hydrodynamic atmosphere, however, $T_{eff}$ is not the dominant
parameter. The hydrodynamic LD correction is instead characterized by
sudden variations due to shocks traveling
through the photosphere. Even when shocks are absent, the LD
correction is still strikingly different from the one predicted by an
hydrostatic model, as it is fully determined by the
time-dependent structure of the expanding/contracting atmosphere. This
dependence from the hydrodynamics results in a larger LD of the
hydrodynamic models with respect to hydrostatic ones.

% -----------------------------------------------------------------------------

\section{Limb darkening Corrections for the VLTI}\label{sec-ld}

The VLTI interferometer, equipped with the near-IR camera AMBER will
be especially suited to measure accurate distances of southern
Cepheids with the geometric Baade-Wesselink method. To obtain
the required accuracy, however, the interferometric measurements
should be accompanied by a detailed modeling of the observed stars,
taking into account the hydrodynamic effects induced by the
pulsations. 

The unprecedented sensitivity of AMBER will result in high accuracy
visibilities ($\sim 10^{-3}$). This will allow the
direct detection of the phase dependence of LD for nearby
Cepheids. Figure~\ref{fig3} shows that the variation of the LD in the
near-IR continuum is of the order of 0.2\% for $\zeta$~Gem. AMBER 
will be able to detect these variations for similar Cepheids at a
distance  of 300-500~pc. The spectral capabilities of AMBER, coupled
with the sensitivity offered by the Unit Telescopes (UT, providing an
expected limiting magnitude K$\sim$14 for $R \sim 1,000$) will
also allow to observe the LD in spectral lines, where the
hydrodynamic effects are much larger. 

The VLTI with AMBER will be the ideal instrument to directly measure
the LD of pulsating Cepheids, and set strong constrains on the
models. This will result in more accurate predictions of the LD
corrections to use for the geometric BW method. The large aperture of
the VLTI telescopes and the large maximum baseline will allow to
detect the pulsation of a large sample of classical
Cepheids. The limiting factor of the interferometers currently
operating in the northern hemisphere is their limiting magnitudes
(e.g. V$\sim$5 for NPOI in the present configuration, and
K$\sim$4.5--5 for the PTI). The 1.8m aperture of the Auxiliary
Telescopes (AT) and the 8m diameter of the UT (necessary to resolve
the spectral lines of Cepheids at $R \sim 1,000$) will extend the
sample of Cepheids that can be observed in the southern hemisphere.
Although the LD will not be directly measured for the most
distant sources with the maximum 200m baseline, the accurate
predictions from the models calibrated with nearby Cepheids will still
allow reliable distance determinations even when the stellar disk is
not fully resolved. This in turn will result in a better calibration of
the PL relation, and thus in a much improved distance scale from
classical Cepheids.

% -----------------------------------------------------------------------------

\acknowledgements
This work was partially supported by NSF grant AST 98-76734. M.K. is a
member of the Chandra Science Center, which is operated under contract
NAS8-39073, and is partially supported by NASA.

% -----------------------------------------------------------------------------
% Bibliography section
% -----------------------------------------------------------------------------

% -----------------------------------------------------------------------------

\end{article}
\end{document}